\documentclass[preprint,pre,aps,amsfonts,amsmath,showpacs,superscriptaddress,nofootinbib, draft]{revtex4}
\usepackage{bm}
\begin{document}
\title{Propagation Effects on the Breakdown of a Linear Amplifier Model: Complex-Mass Schr\"odinger Equation Driven by the Square of a Gaussian Field}
\author{Philippe Mounaix}
\email{mounaix@cpht.polytechnique.fr}
\author{Pierre Collet}
\email{collet@cpht.polytechnique.fr}
\affiliation{Centre de Physique Th\'eorique, UMR 7644 du CNRS, Ecole
Polytechnique, 91128 Palaiseau Cedex, France.}
\author{Joel L. Lebowitz}
\email{lebowitz@math.rutgers.edu}
\affiliation{Departments of Mathematics and Physics, Rutgers, The State
University of New Jersey, Piscataway, New Jersey 08854-8019.}
\date{\today}
\begin{abstract}
Solutions to the equation $\partial_t{\cal E}(x,t)-\frac{i}{2m}\Delta {\cal E}(x,t)=\lambda\vert S(x,t)\vert^2{\cal E}(x,t)$ are investigated, where $S(x,t)$ is a complex Gaussian field with zero mean and specified covariance, and $m\ne 0$ is a complex mass with ${\rm Im}(m)\ge 0$. For real $m$ this equation describes the backscattering of a smoothed laser beam by an optically active medium. Assuming that $S(x,t)$ is the sum of a finite number of independent complex Gaussian random variables, we obtain an expression for the value of $\lambda$ at which the $q$-th moment of $\vert {\cal E}(x,t)\vert$ w.r.t. the Gaussian field $S$ diverges. This value is found to be less or equal for all $m\ne 0$, ${\rm Im}(m)\ge 0$ and $\vert m\vert <+\infty$ than for $\vert m\vert =+\infty$, i.e. when the $\Delta {\cal E}$ term is absent. Our solution is based on a distributional formulation of the Feynman path-integral and the Paley-Wiener theorem. 
\end{abstract}
\pacs{05.10.Gg, 02.50.Ey, 52.38.-r}
\maketitle
%
%%%%%%%%%%%%%%%%%%%%
%
\section{Introduction}\label{sec1}
We investigate the breakdown of linear amplification in a system driven by the square of a Gaussian noise. This problem which models the backscattering of an incoherent laser by an optically active medium was first considered by Akhmanov {\it et al.} in nonlinear optics\ \cite{ADP}, and by Rose and DuBois in laser-plasma interaction\ \cite{RD2}. The latter investigated the divergence of the average solution to the stochastic PDE
\begin{equation}\label{eq1.1}
\left\lbrace\begin{array}{l}
\partial_t{\cal E}(x,t)-\frac{i}{2m}\Delta
{\cal E}(x,t)=
\lambda\vert S(x,t)\vert^2{\cal E}(x,t),\\
t\ge 0,\ x\in\Lambda\subset {\mathbb R}^d,\ {\rm and}\ {\cal E}(x,0)=1,
\end{array}\right.
\end{equation}
heuristically and numerically in the "diffractive case" where ${\rm Im}(m)=0$ and ${\rm Re}(m)\ne 0$. Here $\lambda >0$ is the coupling constant and $S$ is a complex Gaussian noise with zero mean
\footnote{This is the case of interest in laser-plasma interaction and nonlinear optics in which $S$ is the (complex) time-enveloppe of the laser electric field. With the help of some minor modifications, our results carry over straightforwardly to the cases where $S$ is real.}.
More recently, this problem was analyzed from a more rigorous mathematical point of view in\ \cite{ADLM} and\ \cite{ML}. The "diffusive case" in which ${\rm Re}(m)=0$ and ${\rm Im}(m)> 0$ was considered in\ \cite{ADLM}, and the one dimensional diffractive case was considered in\ \cite{ML} for a restrictive class of $S$'s. In the present work we will consider the general case $m\ne 0$ and ${\rm Im}(m)\ge 0$ for $\Lambda$ a $d$-dimensional torus with $d\le 3$. As in\ \cite{ADLM} and\ \cite{ML} we will express the solution to\ (\ref{eq1.1}) formally as the Feynman-Kac path-integral
\begin{equation}\label{eq1.2}
{\cal E}(x,t)=\int_{x(\cdot)\in B(x,t)}
{\rm e}^{\int_0^t\left\lbrack \frac{im}{2}\dot{x}(\tau)^2
+\lambda\vert S(x(\tau),\tau)\vert^2 \right\rbrack\, d\tau}
d\lbrack x(\cdot)\rbrack ,
\end{equation}
where $B(x,t)$ denotes the set of all the continuous paths in $\Lambda$ satisfying $x(t)=x$.
\paragraph*{}In the diffusive case, the right-hand side of\ (\ref{eq1.2}) is just the Wiener integral of $\exp\left(\lambda\int_0^t\vert S(x(\tau),\tau)\vert^2\, d\tau\right)$ over $B(x,t)$. This was used in\ \cite{ADLM} to prove, under some reasonable assumptions on the covariance of $S$, that for every $t>0$ and any positive integer $q$ the average of ${\cal E}(x,t)^q$ over the realizations of $S$, $\langle {\cal E}(x,t)^q\rangle$, diverges as $\lambda$ increases past some critical value smaller (or equal) than in the diffusion-free case (i.e. when $\vert m\vert =+\infty$), with equality holding for a class of $S$. It was conjectured there that this inequality should also apply when diffusion is replaced by diffraction, i.e. $m$ real, $m\ne 0$, the case of physical interest considered by Rose and DuBois in\ \cite{RD2}.
\paragraph*{}The diffractive case is much more difficult because the right-hand side of\ (\ref{eq1.2}) is no longer well defined and one cannot {\it a priori} exclude the possibility that destructive interference between paths makes the sum of divergent contributions finite, raising (possibly to infinity) the critical value of $\lambda$ at which the average of\ (\ref{eq1.2}) diverges. Using heuristics and numerical simulations, Rose and DuBois argued that $\langle\vert {\cal E}(x,t)\vert^2\rangle$ should diverge for every $t>0$ as $\lambda$ increases to some finite critical value\ \cite{RD2}. The conjecture made in\ \cite{ADLM} that diffraction should actually lower the critical coupling (or, at least, not increase it) compared to the case $\vert m\vert =+\infty$ was proved in\ \cite{ML}, for very special choices of $S$, for the divergence of $\langle\vert {\cal E}(x,t)\vert\rangle$.
\paragraph*{}In this paper, we extend the results of\ \cite{ML} to a much wider class of $S$. We analyze the divergence of $\langle\vert {\cal E}(x,t)\vert^q\rangle$ for any positive integer $q$, and we treat both the diffusive and diffractive cases as well as all the intermediate cases between these two limits [i.e. complex $m$ with ${\rm Im}(m)\ge 0$ and $m\ne 0$]. Our strategy for controlling the complex Feynman path-integral\ (\ref{eq1.2}) and determining the critical value of $\lambda$ uses the following three ingredients:
\bigskip

(i) we consider a restricted but quite wide class of $S$ for which ${\cal E}(x,t)$ can be written as a Fourier-Laplace integral w.r.t. a distribution with compact support;
\footnote{The "Fourier-Laplace" integral w.r.t. a distribution with compact support on ${\mathbb R}^N$ is the continuation of the usual Fourier integral from ${\mathbb R}^N$ to ${\mathbb C}^N$.}
\bigskip

(ii) we apply the Paley-Wiener theorem to this Fourier-Laplace integral. This yields the control of\ (\ref{eq1.2}) for "large" $\vert S\vert^2$;
\bigskip

(iii) we average $\vert {\cal E}(x,t)\vert^q$ over the realizations of $S$ and use the control obtained in (ii) to determine the smallest value of $\lambda$ for which this average blows up.
\bigskip

The rest of the paper follows this strategy quite faithfully. In Section\ \ref{sec2} we specify the class of $S$ which we can treat. The distributional formulation of ${\cal E}(x,t)$ is given in Section\ \ref{sec3} and the way to control its growth is explained in Section\ \ref{sec4}. Finally, the determination of the critical value of $\lambda$ and the proof of the conjecture made in\ \cite{ADLM} are given in Section\ \ref{sec5}. It is worth noting that (i) and (ii) do not depend on $S$ being Gaussian and thus apply also in a more general setting.
%
%%%%%%%%%%%%%%%%%%%%
%
\section{Model and definitions}\label{sec2}
We consider the solution to the linear amplifier equation\ (\ref{eq1.1}), written in its integral representation\ (\ref{eq1.2}), with $m$ in $\overline{{\mathbb C}^+}\backslash\lbrace 0\rbrace$, where $\overline{{\mathbb C}^+} \equiv\lbrace m\in {\mathbb C}:\ {\rm Im}(m)\ge 0\rbrace$. We assume that $S$ can be expressed as a finite combination of $M$ complex Gaussian r.v., $s_n$,
\begin{equation}\label{eq2.1}
S(x,t)=\sum_{n=1}^{M}s_n\Phi_n(x,t),
\end{equation}
with
\begin{equation}\label{eq2.2}
\left\lbrace\begin{array}{l}
\langle s_n\rangle =\langle s_n s_m\rangle =0,\\
\langle s_n s_m^\ast\rangle =\delta_{nm}.
\end{array}\right.
\end{equation}
The $\Phi_n$ are normalized such that
$$
\frac{1}{\vert\Lambda\vert}\int_0^1\int_\Lambda \langle\vert S(x,\tau)\vert^2\rangle\, d\tau d^dx =
\frac{1}{\vert\Lambda\vert}\sum_{n=1}^M\int_0^1\int_\Lambda \vert\Phi_n(x,\tau)\vert^2\, d\tau d^dx =
1.
$$
Furthermore, the $\Phi_n(\cdot ,\tau)$ are assumed to have second derivatives bounded uniformly in $\tau\in [0,t]$, and the $\Phi_n(x,\cdot)$ are piecewise continuous for every $x\in\Lambda$ with a finite number of discontinuities in $[0,t]$ for all finite $t$. Note that locally, i.e. for each $x$ and $\tau$, $\vert S(x,\tau)\vert^2$ is a quadratic form of $2M$ real Gaussian r.v. (the real and imaginary parts of the $s_n$), so it is a $\chi^2$ r.v. with $2M$ degrees of freedom.
\paragraph*{}Equation\ (\ref{eq2.1}) generalizes models of spatially smoothed laser beams in which the laser light is represented by a superposition of a finite number of monochromatic beamlets the amplitudes of which are independent r.v.\ \cite{RD1}. For a large number of beamlets these r.v. can be taken as Gaussian and the laser electric field takes on the form\ (\ref{eq2.1}) with $\Phi_n(x,t)\propto\exp\lbrack i(k_n\cdot x+ak_n^2t)\rbrack$, where $k_n$ is the wave vector of the $n$th beamlet and $a>0$ is a (real) constant. It can be checked that all the assumptions made on $S$ are fulfilled.
\paragraph*{}We are interested in the critical coupling $\lambda_{q}(x,t)$ and its Laplacian-free counterpart $\overline{\lambda}_{q}(x,t)$ obtained by setting $m^{-1}=0$ in Eq.\ (\ref{eq1.1}). These quantities are defined by
\begin{subequations}\label{eq2.3}
\begin{eqnarray}
&&\lambda_{q}(x,t)=
\inf\lbrace \lambda>0:
\langle\vert {\cal E}(x,t)\vert^q\rangle =+\infty\rbrace,
\label{eq2.3a}\\
&&\overline{\lambda}_{q}(x,t)=
\inf\lbrace \lambda>0:\langle {\rm e}^{q\lambda\int_0^t
|S(x,\tau)|^2d\tau}\rangle =+\infty\rbrace .
\label{eq2.3b}
\end{eqnarray}
\end{subequations}
Equations\ (\ref{eq2.3}) give the values of $\lambda$ at which $\langle\vert {\cal E}(x,t)\vert^q\rangle$ diverges with and without the Laplacian on the left-hand side of (\ref{eq1.1}). Note that $S$ is not assumed to be homogeneous and the critical coupling will depend on $x$ in general.
%
%%%%%%%%%%%%%%%%%%%%
%
\section{Distributional formulation of $\bm{{\cal E}(x,t;s)}$}\label{sec3}
Let $s$ be the $M$-dimensional Gaussian random vector the elements of which are the $s_n$, and $\gamma (x,\tau)$ the $M\times M$ Hermitian matrix defined by
$$
\gamma_{nm}(x,\tau)=\Phi_n^\ast(x,\tau)\Phi_m(x,\tau).
$$
Inserting (\ref{eq2.1}) into the right-hand side of (\ref{eq1.2}) yields 
\begin{equation}\label{eq3.1}
{\cal E}(x,t;s)=\int_{x(\cdot)\in B(x,t)}
{\rm e}^{\int_0^t\left\lbrack \frac{im}{2}\dot{x}(\tau)^2
+\lambda\, s^{\dag}\gamma (x(\tau),\tau)s \right\rbrack\, d\tau}
d\lbrack x(\cdot)\rbrack ,
\end{equation}
where we have made the dependence of ${\cal E}(x,t)$ on the realization of $s$ explicit. In order to make\ (\ref{eq3.1}) more appropriate to a distributional formulation it is desirable to replace the quadratic form $s^{\dag}\gamma (x(\tau),\tau)s$ with its monomial decomposition. One obtains
\begin{equation}\label{eq3.2}
{\cal E}(x,t;s)=\int_{x(\cdot)\in B(x,t)}{\rm e}^{\frac{im}{2}\int_0^t\dot{x}(\tau)^2d\tau}
\exp\left\lbrack
\lambda\sum_{i=1}^N k_i(s)\int_0^t\varphi_i(x(\tau),\tau)\, d\tau
\right\rbrack\,  d\lbrack x(\cdot)\rbrack ,
\end{equation}
with $N=M^2$, and where the $\varphi_i$ are $N$ real valued functions given by $\gamma_{nn}$, $\sqrt{2}{\rm Re}(\gamma_{nm})$, and $\sqrt{2}{\rm Im}(\gamma_{nm})$, $n<m$. The components of the vector $k(s)\in {\mathbb R}^N$ are given by $\vert s_n\vert^2$, $\sqrt{2}{\rm Re}(s_n s_m^\ast)$, and $\sqrt{2}{\rm Im}(s_n s_m^\ast)$, $n<m$. It can be checked that
\begin{equation}\label{eq3.3}
\| k(s)\| =\| s\|^2,
\end{equation}
where $\| k(s)\| =\left(\sum_{i=1}^Nk_i(s)^2\right)^{1/2}$ and $\| s\| =\left(\sum_{i=1}^M\vert s_i\vert^2\right)^{1/2}$. We first give a heuristic derivation of the distributional formulation of\ (\ref{eq3.1}). Then, we set it on a much firmer ground by justifying it rigorously from a mathematical point of view.
%
%\bigskip
\subsection{Heuristics}
Inserting the identity
$$
1=\prod_{i=1}^N\int_{{\mathbb R}}\delta\left(u_i-\int_0^t\varphi_i(x(\tau),\tau)\,
d\tau\right)\, du_i,
$$
in the path-integral\ (\ref{eq3.2}) and permuting the path- and $u$-integrals, one obtains
\begin{equation}\label{eq3.4}
{\cal E}(x,t;s)=\int\cdots\int_{{\mathbb R}^N}
G_{x,t}(u)\, {\rm e}^{\lambda k(s)\cdot u}\prod_{i=1}^N du_i,
\end{equation}
with
\begin{equation}\label{eq3.5}
G_{x,t}(u_1,...u_N)=
\int_{x(\cdot)\in B(x,t)}
{\rm e}^{\frac{im}{2}\int_0^t\dot{x}(\tau)^2d\tau}
\prod_{i=1}^N\delta\left(u_i-\int_0^t\varphi_i(x(\tau),\tau)\,
d\tau\right)\, d\lbrack x(\cdot)\rbrack .
\end{equation}
As a Feynman-Kac path-integral, the expression\ (\ref{eq3.5}) is not well defined.  A possible way to make it meaningful consists in writing $G_{x,t}$ as the Fourier transform w.r.t. $\eta$ of some function $\Psi(x,t;\eta)$:
\begin{equation}\label{eq3.6}
G_{x,t}(u)=\frac{1}{(2\pi)^N}\int\cdots\int_{{\mathbb R}^N} \Psi(x,t;\eta)\, {\rm e}^{iu\cdot\eta}
\prod_{i=1}^N d\eta_i ,
\end{equation}
in which $\Psi(x,t;\eta)$ has a well defined meaning. Fourier transforming\ (\ref{eq3.5}) w.r.t. $u$ and permuting the path- and $u$-integrals, one obtains
\begin{equation}\label{eq3.7}
\Psi(x,t;\eta)=
\int_{x(\cdot)\in B(x,t)}
{\rm e}^{i\int_0^t\left\lbrack
\frac{m}{2}\dot{x}(\tau)^2-
V(x(\tau),\tau;\eta)\right\rbrack\, d\tau} d\lbrack x(\cdot)\rbrack ,
\end{equation}
where $V(x,t;\eta)$ is given by
\begin{equation}\label{eq3.8}
V(x,t;\eta)\equiv\sum_{i=1}^N\eta_i\varphi_i(x,t).
\end{equation}
We now observe that Eq.\ (\ref{eq3.7}) is the path-integral solution to the Schr\"odinger equation
\begin{equation}\label{eq3.9}
\left\lbrace\begin{array}{l}
i\partial_t\Psi(x,t;\eta)=-\frac{1}{2m}\Delta\Psi(x,t;\eta)
+V(x,t;\eta)\Psi(x,t;\eta),\\
t\ge 0,\ x\in\Lambda,\ {\rm and}\ \Psi(x,0;\eta)=1.
\end{array}\right. 
\end{equation}
This yields a well defined $\Psi(x,t;\eta)$.
\paragraph*{}The permutation of path- and ordinary integrals, as well as the formal Feynman-Kac path-integral used in the derivation of Eqs.\ (\ref{eq3.4}),\ (\ref{eq3.6}), and\ (\ref{eq3.9}) above require justification. The work by Cartier and DeWitt-Morette\ \cite{CDM} suggests that we {\it define} the path-integral\ (\ref{eq3.1}) by the right-hand side of\ (\ref{eq3.4}) in which $G_{x,t}$ is defined by its Fourier transform given as the solution to Eq.\ (\ref{eq3.9}). We now prove the validity of this approach.
%
%\bigskip
\subsection{The distributional formulation}
Let $\Psi(x,t;\eta)$ be the solution to\ (\ref{eq3.9}) where $V(x,t;\eta)$ is given by\ (\ref{eq3.8}) with $\eta\in {\mathbb C}^N$. Let $a_i=\inf_{x(\cdot)\in B(x,t)}\int_0^t\varphi_i(x(\tau),\tau)\, d\tau$ and $b_i=\sup_{x(\cdot)\in B(x,t)}\int_0^t\varphi_i(x(\tau),\tau)\, d\tau$. Then the following lemma holds.
\paragraph*{}
\bigskip
\noindent {\bf Lemma 1.}
{\it For every $t>0$, $x\in\Lambda$, and $m\in\overline{{\mathbb C}^+}\backslash\lbrace 0\rbrace$,
\paragraph*{}{\it (i)} $G_{x,t}$ defined by\ (\ref{eq3.6}) is a distribution with compact support on ${\mathbb R}^N$ and ${\rm supp}G_{x,t}\subset\lbrack a_1,b_1\rbrack\times ...\times\lbrack a_N,b_N\rbrack$;
\paragraph*{}{\it (ii)} ${\cal E}(x,t;s)$ defined by\ (\ref{eq3.4}) is the solution to\ (\ref{eq1.1}).}
\paragraph*{}
\bigskip
\noindent {\it Proof.}
Taking the derivative of\ (\ref{eq3.9}) with respect to $\eta_i^\ast$, the complex conjugate of $\eta_i$, and using $\partial_{\eta_i^\ast}V(x,t;\eta)=0$ which follows from analyticity of $V(x,t;\eta)$ in $\eta$ [see Eq.\ (\ref{eq3.8})], one finds that $\partial_{\eta_i^\ast}\Psi(x,t;\eta)$ evolves in time according to the same equation\ (\ref{eq3.9}) with the initial condition $\partial_{\eta_i^\ast}\Psi(x,0;\eta)=0$. Thus, $\partial_{\eta_i^\ast}\Psi(x,t;\eta)=0$ for all $t\ge 0$ and $\eta\in {\mathbb C}^N$ which implies that $\Psi(x,t;\eta)$ is analytic in $\eta$.
\paragraph*{}Let $\tilde{\Psi}(x,t;\eta)=\Psi(x,t;\eta)\exp(-it\sum_{i=1}^N\eta_ic_i)$ where the constants $c_i\in {\mathbb R}$ will be specified later. $\tilde{\Psi}$ is the solution to\ (\ref{eq3.9}) with $V$ given by\ (\ref{eq3.8}) in which the $\varphi_i$ are replaced by $\tilde{\varphi}_i=\varphi_i+c_i$. Let $\epsilon_i={\rm sgn}\lbrack {\rm Im}(\eta_i)\rbrack$. From the equation\ (\ref{eq3.9}) for $\tilde{\Psi}$ and the Schwartz inequality one obtains
\begin{eqnarray}\label{eq3.10a}
\frac{d}{dt}\|\tilde{\Psi}\|_2^2&=&
-\frac{{\rm Im}(m)}{\vert m\vert^2}\|\nabla\tilde{\Psi}\|_2^2
+2\sum_{i=1}^N {\rm Im}(\eta_i)\int_\Lambda\tilde{\varphi}_i\vert\tilde{\Psi}\vert^2d^dx \nonumber \\
&\le& 2\|\tilde{\Psi}\|_2^2\sum_{i=1}^N \vert{\rm Im}(\eta_i)\vert\sup_{x\in\Lambda}
(\epsilon_i\tilde{\varphi}_i),
\end{eqnarray}
\begin{eqnarray}\label{eq3.10b}
\frac{d}{dt}\|\nabla\tilde{\Psi}\|_2^2&=&
-\frac{{\rm Im}(m)}{\vert m\vert^2}\|\Delta\tilde{\Psi}\|_2^2
+2\sum_{i=1}^N {\rm Im}(\eta_i)\int_\Lambda\tilde{\varphi}_i
\vert\nabla\tilde{\Psi}\vert^2d^dx \nonumber \\
&+&2{\rm Im}\sum_{i=1}^N\eta_i\int_\Lambda\left\lbrack\tilde{\Psi}
\nabla\tilde{\Psi}^{\ast}\right\rbrack\cdot\nabla\tilde{\varphi}_i\, d^dx \\
&\le& 2\|\nabla\tilde{\Psi}\|_2^2\sum_{i=1}^N
\vert{\rm Im}(\eta_i)\vert\sup_{x\in\Lambda}(\epsilon_i\tilde{\varphi}_i)
+2\|\nabla\tilde{\Psi}\|_2\|\tilde{\Psi}\|_2
\sum_{i=1}^N\vert\eta_i\vert\|\nabla\tilde{\varphi}_i\|_{\infty}, \nonumber
\end{eqnarray}
and
\begin{eqnarray}
\frac{d}{dt}\|\Delta\tilde{\Psi}\|_2^2&=&
-\frac{{\rm Im}(m)}{\vert m\vert^2}\|\nabla\Delta\tilde{\Psi}\|_2^2
+2\sum_{i=1}^N {\rm Im}(\eta_i)\int_\Lambda\tilde{\varphi}_i
\vert\Delta\tilde{\Psi}\vert^2d^dx \nonumber \\
&+&2{\rm Im}\sum_{i=1}^N\eta_i\int_\Lambda \Delta\tilde{\Psi}^{\ast}\left\lbrack
\tilde{\Psi}\Delta\tilde{\varphi}_i+2\nabla\tilde{\Psi}\cdot\nabla\tilde{\varphi}_i
\right\rbrack\, d^dx \label{eq3.10c} \\
&\le& 2\|\Delta\tilde{\Psi}\|_2^2\sum_{i=1}^N
\vert{\rm Im}(\eta_i)\vert\sup_{x\in\Lambda}(\epsilon_i\tilde{\varphi}_i) \nonumber \\
&+& 2\|\Delta\tilde{\Psi}\|_2\sum_{i=1}^N\vert\eta_i\vert
\left(\|\tilde{\Psi}\|_2\|\Delta\tilde{\varphi}_i\|_{\infty}
+2\|\nabla\tilde{\Psi}\|_2\|\nabla\tilde{\varphi}_i\|_{\infty}
\right),\nonumber
\end{eqnarray}
where $\|\cdot\|_2$ and $\|\cdot\|_{\infty}$ respectively denote the $L^2$ and uniform norms
\footnote{In the case of a vector field $v(x,t)\in {\mathbb C}^d$, these norms are to be understood as $\| v\|_2(t)=\left(\int_\Lambda v(x,t)\cdot v(x,t)^\ast d^dx\right)^{1/2}$ and $\| v\|_{\infty}(t)=\sup_{x\in\Lambda}\left(\sqrt{v(x,t)\cdot v(x,t)^\ast}\right)$.}
on $\Lambda$ for given $t$ and $\eta$. Both $\|\nabla\tilde{\varphi}_i\|_{\infty}(t)$ and $\|\Delta\tilde{\varphi}_i\|_{\infty}(t)$ are bounded by assumption. Integrating then the inequality\ (\ref{eq3.10a}) over time from $0$ to $t$, one obtains
\begin{equation}\label{eq3.11}
\|\tilde{\Psi}\|_2(t,\eta) \le\vert\Lambda\vert^{1/2}
{\rm e}^{\sum_{i=1}^N \vert{\rm Im}(\eta_i)\vert
\int_0^t\sup_{x\in\Lambda}
\lbrack\epsilon_i\tilde{\varphi}_i(x,\tau)\rbrack\, d\tau}.
\end{equation}
Similarly, by integrating\ (\ref{eq3.10b}) and\ (\ref{eq3.10c}) one finds
\begin{equation}\label{eq3.12}
\|\Delta\tilde{\Psi}\|_2(t,\eta)\le\left\lbrack C_1t\sum_{i=1}^N\vert\eta_i\vert
+C_2t^2\left(\sum_{i=1}^N\vert\eta_i\vert\right)^2\right\rbrack
{\rm e}^{\sum_{i=1}^N \vert{\rm Im}(\eta_i)\vert
\int_0^t\sup_{x\in\Lambda}
\lbrack\epsilon_i\tilde{\varphi}_i(x,\tau)\rbrack\, d\tau},
\end{equation}
where $C_1$ and $C_2$ are finite and independent of $\eta$ and $m$. We now substitute\ (\ref{eq3.11}) and\ (\ref{eq3.12}) into the right-hand side of the Sobolev-type inequality below, valid for $d\le 3$ (see e.g.\ \cite{Thirring} pp 106-107),
$$
\vert\tilde{\Psi}(x,t;\eta)\vert\le C_3
\left\lbrack\|\tilde{\Psi}\|_2(t,\eta) +
\|\Delta\tilde{\Psi}\|_2(t,\eta)\right\rbrack ,
$$
with $C_3$ finite and independent of $\eta$ and $m$. This yields
\begin{equation}\label{eq3.13}
\vert\tilde{\Psi}(x,t;\eta)\vert\le\left\lbrack A+Bt\sum_{i=1}^N\vert\eta_i\vert
+Ct^2\left(\sum_{i=1}^N\vert\eta_i\vert\right)^2\right\rbrack
{\rm e}^{\sum_{i=1}^N \vert{\rm Im}(\eta_i)\vert
\int_0^t\sup_{y\in\Lambda}
\lbrack\epsilon_i\tilde{\varphi}_i(y,\tau)\rbrack\, d\tau},
\end{equation}
where $A$, $B$, and $C$ are finite and independent of $\eta$ and $m$. Take
$$
c_i=-\frac{1}{2t}\left\lbrace
\int_0^t\sup_{x\in\Lambda}\lbrack\varphi_i(x,\tau)\rbrack\, d\tau
+\int_0^t\inf_{x\in\Lambda}\lbrack\varphi_i(x,\tau)\rbrack\, d\tau
\right\rbrace ,
$$
and define
$$
\kappa_i\equiv\int_0^t\sup_{x\in\Lambda}
\lbrack\epsilon_i\tilde{\varphi}_i(x,\tau)\rbrack\, d\tau
=\frac{1}{2}\left\lbrace
\int_0^t\sup_{x\in\Lambda}\lbrack\varphi_i(x,\tau)\rbrack\, d\tau
-\int_0^t\inf_{x\in\Lambda}\lbrack\varphi_i(x,\tau)\rbrack\, d\tau
\right\rbrace .
$$
Note that, with this choice of $c_i$,  $\kappa_i$ is independent of $\epsilon_i$. Since $\kappa_i\ge 0$, one can bound the right side of\ (\ref{eq3.13}) by
\begin{equation}\label{eq3.14}
\vert\tilde{\Psi}(x,t;\eta)\vert\le\left\lbrack A+Bt\sum_{i=1}^N\vert\eta_i\vert
+Ct^2\left(\sum_{i=1}^N\vert\eta_i\vert\right)^2\right\rbrack
{\rm e}^{\sum_{i=1}^N \kappa_i\vert\eta_i\vert}.
\end{equation}
\paragraph*{}Let $\tilde{G}_{x,t}$ be defined by Eq.\ (\ref{eq3.6}) in which $\Psi$ is replaced by $\tilde{\Psi}$. By the Paley-Wiener theorem in the formulation given in\ \cite{PWS} (Theorem XVI in Chapter VII), it follows from the analyticity of $\tilde{\Psi}(x,t;\eta)$ in $\eta$ and Eq.\ (\ref{eq3.14}) that $\tilde{G}_{x,t}$ is a distribution with compact support on ${\mathbb R}^N$ and ${\rm supp}\tilde{G}_{x,t}\subset\left\lbrack -\kappa_1 ,\kappa_1\right\rbrack\times ...\times\left\lbrack -\kappa_N ,\kappa_N\right\rbrack$. From the definition of $\tilde{\Psi}$ one has $\tilde{G}_{x,t}(u)=G_{x,t}(u-ct)$, which implies immediately by translation that $G_{x,t}$ is also a distribution with compact support on ${\mathbb R}^N$ and ${\rm supp}G_{x,t}\subset\left\lbrack\alpha_1 ,\beta_1\right\rbrack\times ...\times\left\lbrack\alpha_N ,\beta_N\right\rbrack$, where $\alpha_i=\int_0^t\inf_{x\in\Lambda}\varphi_i(\tau)\, d\tau$ and $\beta_i=\int_0^t\sup_{x\in\Lambda}\varphi_i(\tau)\, d\tau$. The permutation of time integral and space supremum (resp. infimum) can then be performed by using Lemma A1 with $W=\pm\varphi_i$ (see Appendix\ \ref{appA}). One obtains $\alpha_i=a_i$ and $\beta_i=b_i$, yielding
\begin{equation}\label{eq3.15}
{\rm supp}G_{x,t}\subset
\left\lbrack a_1 ,b_1\right\rbrack
\times ...\times
\left\lbrack a_N ,b_N\right\rbrack .
\end{equation}
It is worth noting that, heuristically,\ (\ref{eq3.15}) follows immediately from the formal expression\ (\ref{eq3.5}) since the product of the delta functions vanishes identically outside $\lbrack a_1,b_1\rbrack\times ...\times\lbrack a_N,b_N\rbrack$.
\paragraph*{}It remains to prove that ${\cal E}(x,t;s)$ defined by the r.h.s. of\ (\ref{eq3.4}) is the solution to\ (\ref{eq1.1}). To this end it suffices to note that Eq.\ (\ref{eq3.4}) can be written as ${\cal E}(x,t;s)=\Psi(x,t;\eta =i\lambda k(s))$ which is the solution to\ (\ref{eq3.9}) with $\eta =i\lambda k(s)$ in the potential\ (\ref{eq3.8}). It can be checked that the latter equation is indeed Equation\ (\ref{eq1.1}) [reconstruct $s^\dag\gamma s$ from its monomial decomposition and multiply\ (\ref{eq3.9}) by $-i$], which completes the proof of Lemma 1.
%
%%%%%%%%%%%%%%%%%%%%
%
\section{Controlling the growth of $\bm{\vert{\cal E}(x,t;s)\vert^q}$}\label{sec4}
The advantage gained by recasting\ (\ref{eq3.1}) as\ (\ref{eq3.4}) is that the latter formulation is suitable for a straightforward application of the Paley-Wiener theorem (see e.g. \ \cite{PWS} Theorem XVI in Chapter VII, and\ \cite{PW} Theorem 7.4 in Chapter VI), offering the possibility of controlling the growth of ${\cal E}(x,t;s)$ as $\| s\|\rightarrow +\infty$. This is embodied in Lemma 2 below. Let $\hat{s}\equiv s/\| s\|$ be the direction of $s$ in ${\mathbb C}^M$ and $H_{x,t}(\hat{s})=\sup_{x(\cdot)\in B(x,t)}\int_0^t U(x(\tau),\tau;\hat{s})\, d\tau$, with $U(x,\tau;\hat{s})=\sum_{i=1}^N\hat{k}(s)_i\varphi_i(x,\tau)$ where $\hat{k}(s)= k(s)/\| k(s)\|$.
\paragraph*{}
\bigskip
\noindent {\bf Lemma 2.}
{\it For every $t>0$, $x\in\Lambda$, $m\in\overline{{\mathbb C}^+}\backslash\lbrace 0\rbrace$, and $q$ a positive integer, one has
\begin{equation}\label{eq4.1}
\limsup_{\| s\|\rightarrow +\infty}\frac{\ln\left\vert {\cal E}(x,t;s)\right\vert^q}
{\| s\|^2}= q\lambda H_{x,t}(\hat{s}),
\end{equation}
along every given direction $\hat{s}$ in ${\mathbb C}^M$.}
\paragraph*{}
\bigskip
\noindent {\it Proof.}
From Eqs.\ (\ref{eq3.3}) and\ (\ref{eq3.4}), one can rewrite the left-hand side of\ (\ref{eq4.1}) as
$$
\limsup_{\| s\|\rightarrow +\infty}\frac{\ln\left\vert {\cal E}(x,t;s)\right\vert^q}
{\| s\|^2}=q\lambda\limsup_{\| k(s)\|\rightarrow +\infty}
\frac{1}{\lambda\| k(s)\|}\ln\left\vert\int\cdots\int_{{\mathbb R}^N}
G_{x,t}(u)\, {\rm e}^{\lambda k(s)\cdot u}\prod_{i=1}^Ndu_i \right\vert .
$$
Fixing the direction of $s$ in ${\mathbb C}^M$ also fixes the direction of $k(s)$ in ${\mathbb R}^N$. Write $u=v\hat{k}(s)+u_{\perp}$, with $u_{\perp}\cdot\hat{k}(s)=0$, replace $G_{x,t}(u)$ by its Fourier representation\ (\ref{eq3.6}), and let $\eta_{\vert\vert}$ and $\eta_{\perp}$ denote the Fourier conjugated variables of $v$ and $u_{\perp}$, respectively. One obtains,
\begin{eqnarray*}
&&\int\cdots\int_{{\mathbb R}^N}
G_{x,t}(u)\, {\rm e}^{\lambda k(s)\cdot u}\prod_{i=1}^Ndu_i = \\
&&\int\cdots\int_{{\mathbb R}^N}
{\rm e}^{\lambda \| k(s)\| v}\left\lbrack
\frac{1}{(2\pi)^N}\int\cdots\int_{{\mathbb R}^N} \Psi(x,t;\eta)\, 
{\rm e}^{i(v\eta_{\vert\vert}+u_{\perp}\cdot\eta_{\perp})}
d\eta_{\vert\vert}\prod_{i=1}^{N-1} d\eta_{\perp i}\right\rbrack\, 
dv\prod_{i=1}^{N-1}du_{\perp i}.
\end{eqnarray*}
Performing the integration over $u_{\perp}$ first, and then the one over $\eta_{\perp}$, one finds that the latter expression reduces to
$$
\int\cdots\int_{{\mathbb R}^N}
G_{x,t}(u)\, {\rm e}^{\lambda k(s)\cdot u}\prod_{i=1}^Ndu_i =
\int_{{\mathbb R}} g_{x,t}(v)\, {\rm e}^{\lambda \| k(s)\| v}dv,
$$
with
$$
g_{x,t}(v)=\frac{1}{(2\pi)}\int_{{\mathbb R}}\Psi(x,t;\eta_{\vert\vert},\eta_{\perp}=0)\, 
{\rm e}^{iv\eta_{\vert\vert}}d\eta_{\vert\vert},
$$
where $\Psi(x,t;\eta_{\vert\vert},\eta_{\perp}=0)$ is the solution to\ (\ref{eq3.9}) with $V(x,t;\eta)=\eta_{\vert\vert}U(x,t;\hat{s})$. Let $a=\inf_{x(\cdot)\in B(x,t)}\int_0^tU(x(\tau),\tau;\hat{s})\, d\tau$ and $b=\sup_{x(\cdot)\in B(x,t)}\int_0^tU(x(\tau),\tau;\hat{s})\, d\tau =H_{x,t}(\hat{s})$. By Lemma 1, $g_{x,t}$ is a distribution with compact support on ${\mathbb R}$ and ${\rm supp}g_{x,t}\subset\lbrack a,b\rbrack$. This implies that $\sup\lbrace v:v\in{\rm supp}g_{x,t}\rbrace\le H_{x,t}(\hat{s})$, and by the Paley-Wiener theorem,
\begin{equation}\label{eq4.2}
\limsup_{\| k(s)\|\rightarrow +\infty}
\frac{1}{\lambda\| k(s)\|}\ln\left\vert\int_{{\mathbb R}} g_{x,t}(v)\, {\rm e}^{\lambda \| k(s)\| v}dv\right\vert
\le H_{x,t}(\hat{s}).
\end{equation}
We now prove that\ (\ref{eq4.2}) is an equality. Suppose that $\exists\varepsilon >0$ such that
\begin{equation}\label{eq4.3}
\limsup_{\| k(s)\|\rightarrow +\infty}
\frac{1}{\lambda\| k(s)\|}\ln\left\vert
\int_{{\mathbb R}} g_{x,t}(v)\, {\rm e}^{\lambda \| k(s)\| v}dv
\right\vert\le H_{x,t}(\hat{s})-\varepsilon .
\end{equation}
Then, according to the Paley-Wiener theorem, $\sup\lbrace v:v\in {\rm supp}g_{x,t}\rbrace\le H_{x,t}(\hat{s})-\varepsilon$. It is shown in Appendix\ \ref{appB} that $\sup\lbrace v:v\in {\rm supp}g_{x,t}\rbrace =H_{x,t}(\hat{s})$, yielding $H_{x,t}(\hat{s})\le H_{x,t}(\hat{s})-\varepsilon$, in contradiction with $\varepsilon >0$. Thus, Eq.\ (\ref{eq4.3}) is false and one obtains
$$
\limsup_{\| k(s)\|\rightarrow +\infty}
\frac{1}{\lambda\| k(s)\|}\ln
\left\vert\int_{{\mathbb R}} g_{x,t}(v)\, {\rm e}^{\lambda \| k(s)\| v}dv\right\vert
=H_{x,t}(\hat{s}).
$$
This completes the proof of Lemma 2.
%
%%%%%%%%%%%%%%%%%%%%
%
\section{Determination of $\bm{\lambda_{q}(x,t)}$ and comparison to
$\bm{\overline{\lambda}_{q}(x,t)}$}\label{sec5}
In this section we prove the conjecture made in Ref.\ \cite{ADLM} that $\lambda_q \le\overline{\lambda}_q$, in the case where $S(x,t)$ is given by\ (\ref{eq2.1}). Since we wish to express the results in terms of eigenvalues of the correlation function $\langle S^\ast (x(t),t)S(x(t'),t')\rangle$, we begin with a technical preliminary linking these eigenvalues to those of the matrix $\int_0^t\gamma(x(\tau),\tau)\, d\tau$.
\paragraph*{}Let $\mu_1\lbrack x(\cdot)\rbrack\ge\mu_2\lbrack x(\cdot)\rbrack\ge\cdots\ge 0$
be the eigenvalues of the covariance operator $T_{x(\cdot)}$ acting on
$f(\tau)\in L^2(d\tau)$, defined by
\begin{equation}\label{eq5.1}
(T_{x(\cdot)} f)(\tau)=\int_0^t \langle S^\ast (x(\tau),\tau)S(x(\tau'),\tau')\rangle f(\tau')\, d\tau',
\end{equation}
with $0\le\tau,\tau'\le t$ and $x(\cdot)\in B(x,t)$. Let $f_i(\tau)\in L^2(d\tau)$ be the eigenfunction associated with $\mu_i\lbrack x(\cdot)\rbrack$ and define the vector $\sigma_i\in {\mathbb C}^M$ by $\sigma_{in}=\int_0^t\Phi_n^\ast(x(\tau),\tau)f_i^\ast(\tau)\, d\tau$. From\ (\ref{eq5.1}),\ (\ref{eq2.1}), and\ (\ref{eq2.2}) one has
\begin{equation}\label{eq5.2}
(T_{x(\cdot)} f_i)(\tau)=\sum_{m=1}^M\Phi_m^\ast(x(\tau),\tau)\sigma_{im}^\ast
=\mu_i\lbrack x(\cdot)\rbrack f_i(\tau),
\end{equation}
and
\begin{eqnarray}
\mu_i\lbrack x(\cdot)\rbrack\sigma_{in}&=&
\int_0^t\Phi_n^\ast(x(\tau),\tau)(T_{x(\cdot)} f_i)^\ast(\tau)\, d\tau \nonumber \\
&=&\sum_{m=1}^M\sigma_{im}
\int_0^t\Phi_n^\ast(x(\tau),\tau)\Phi_m(x(\tau),\tau)\, d\tau \label{eq5.3} \\
&=&\sum_{m=1}^M\left\lbrack\int_0^t\gamma_{nm}(x(\tau),\tau)\, d\tau\right\rbrack\sigma_{im},\nonumber
\end{eqnarray}
It follows from the last equality of\ (\ref{eq5.3}) that any non vanishing eigenvalue of $T_{x(\cdot)}$ is also an eigenvalue of $\int_0^t\gamma(x(\tau),\tau)\, d\tau$ with eigenvector $\sigma_i$. Conversely, any non vanishing eigenvalue of $\int_0^t\gamma(x(\tau),\tau)\, d\tau$ is also eigenvalue of $T_{x(\cdot)}$ with eigenfunction $f_i(\tau)=\mu_i^{-1}\sum_{m=1}^M\sigma_{im}^\ast\Phi_m^\ast(x(\tau),\tau)$ [see Eq.\ (\ref{eq5.2})]. Thus, there is a one-to-one relationship between the non vanishing eigenvalues of $T_{x(\cdot)}$ and $\int_0^t\gamma(x(\tau),\tau)\, d\tau$. In the sequel $\mu_1\lbrack x(\cdot)\rbrack$ will denote the largest eigenvalue of $T_{x(\cdot)}$ and of $\int_0^t\gamma(x(\tau),\tau)\, d\tau$. Define
$$
\mu_{x,t}=\sup_{x(\cdot)\in B(x,t)}\mu_1\lbrack x(\cdot)\rbrack .
$$
One can now prove the following proposition:
\paragraph*{}
\bigskip
\noindent {\bf Proposition.}
{\it For every $t>0$ and $x\in\Lambda$,
$\lambda_{q}(x,t)=(q\mu_{x,t})^{-1}\le\overline{\lambda}_{q}(x,t)$.}
\paragraph*{}
\bigskip
\noindent {\it Proof.}
First we prove $\lambda_{q}(x,t)\ge (q\mu_{x,t})^{-1}$. Expressing $U(x(\tau),\tau;\hat{s})$ in terms of the quadratic form $s^{\dag}\gamma (x(\tau),\tau)s$ in the expression for $H_{x,t}(\hat{s})$, one has
$$
H_{x,t}(\hat{s})=\sup_{x(\cdot)\in B(x,t)}\, \frac{s^\dag}{\vert\vert s\vert\vert}
\left\lbrack\int_0^t\gamma(x(\tau),\tau)\, d\tau\right\rbrack
\frac{s}{\vert\vert s\vert\vert} \le
\mu_{x,t} .
$$
Hence, by Lemma 2,
$$
\limsup_{\vert\vert s\vert\vert\rightarrow +\infty}\frac{\ln\left\vert {\cal E}(x,t;s)\right\vert^q}
{\vert\vert s\vert\vert^2}\le q\lambda\mu_{x,t} .
$$
This implies that for every $\lambda <(q\mu_{x,t})^{-1}$,
\begin{equation}\label{eq5.4}
\langle\vert {\cal E}(x,t)\vert^q\rangle =
\int\cdots\int_{{\mathbb C}^M} {\rm e}^{-\vert\vert s\vert\vert^2}
\vert {\cal E}(x,t;s)\vert^q\prod_{n=1}^M \frac{d^2s_n}{\pi}<+\infty ,
\end{equation}
which proves $\lambda_{q}(x,t)\ge (q\mu_{x,t})^{-1}$.
\paragraph*{}We now prove the inequality $\lambda_{q}(x,t)\le (q\mu_{x,t})^{-1}$ [or, more exactly, $\lambda_{q}(x,t)\le (q\mu_{x,t} -0^+)^{-1}$]. To this end we follow the same line of reasoning as in Ref.\ \cite{ML}. Let ${\cal A}(r)=\lbrace z\in{\mathbb C}^M:\vert z_n\vert\le r,\ 1\le n\le M \rbrace$. For Eq.\ (\ref{eq5.4}) to hold it is necessary that, for every $r >0$,
\begin{equation}\label{eq5.5}
\lim_{\vert\vert s\vert\vert\rightarrow +\infty}
\int\cdots\int_{{\cal A}(r)} {\rm e}^{-\vert\vert s+s'\vert\vert^2}
\vert {\cal E}(x,t;s+s')\vert^q\prod_{n=1}^M \frac{d^2s'_n}{\pi}=0,
\end{equation}
along every direction $\hat{s}$ in ${\mathbb C}^M$. For any fixed $s\in{\mathbb C}^M$, ${\rm e}^{-2s^\ast\cdot z/q}{\cal E}(x,t;s+z)$ is an entire function of $z\in{\mathbb C}^M$ and hence $\vert{\rm e}^{-2s^\ast\cdot z/q}{\cal E}(x,t;s+z)\vert^q$ is subharmonic w.r.t. each component of $z$\ \cite{Hay}. Thus, writing
$$
{\rm e}^{-\vert\vert s+s'\vert\vert^2}={\rm e}^{-\vert\vert s\vert\vert^2}{\rm e}^{-\vert\vert s'\vert\vert^2}\left\vert\exp\left(-\frac{2}{q}s^\ast\cdot s'\right)\right\vert^q,
$$
in the integral\ (\ref{eq5.5}), one obtains by the subharmonicity
\begin{eqnarray*}
&&\int\cdots\int_{{\cal A}(r)}{\rm e}^{-\vert\vert s+s'\vert\vert^2}\vert{\cal E}(x,t;s+s')\vert^q
\prod_{n=1}^M\frac{d^2s'_n}{\pi} \\
&&={\rm e}^{-\vert\vert s\vert\vert^2}
\int\cdots\int_{{\cal A}(r)}{\rm e}^{-\vert\vert s'\vert\vert^2}
\left\vert {\rm e}^{-\frac{2}{q}s^\ast\cdot s'}{\cal E}(x,t;s+s')\right\vert^q
\prod_{n=1}^M\frac{d^2s'_n}{\pi} \\
&&\ge\left\lbrack 1-\exp\left(-r^2\right)\right\rbrack^M
{\rm e}^{-\vert\vert s\vert\vert^2}\vert {\cal E}(x,t;s)\vert^q,
\end{eqnarray*}
and the condition\ (\ref{eq5.5}) implies
\begin{equation}\label{eq5.6}
\lim_{\vert\vert s\vert\vert\rightarrow +\infty}{\rm e}^{-\vert\vert s\vert\vert^2}
\vert {\cal E}(x,t;s)\vert^q =0,
\end{equation}
along every direction $\hat{s}$ in ${\mathbb C}^M$. Since every element of the matrix $\int_0^t\gamma(x(\tau),\tau)\, d\tau$ is a continuous functional of $x(\cdot)\in B(x,t)$ with the uniform norm on $\lbrack 0,t\rbrack$ (see the appendix\ \ref{appB}), its eigenvalues are also continuous functionals of $x(\cdot)$. Accordingly, $\forall\varepsilon >0$
$\exists x_{\varepsilon}(\cdot)\in B(x,t)$ such that $\mu_{x,t} -\varepsilon\le\mu_1\lbrack x_{\varepsilon}(\cdot)\rbrack\le\mu_{x,t}$. Let $\sigma_{\varepsilon}\in {\mathbb C}^M$ (with $\vert\vert \sigma_{\varepsilon}\vert\vert =1$) be an eigenvector of $\int_0^t\gamma(x_{\varepsilon}(\tau),\tau)\, d\tau$ associated with the eigenvalue $\mu_1\lbrack x_{\varepsilon}(\cdot)\rbrack$ and take $\hat{s} =\sigma_{\varepsilon}$, then
\begin{eqnarray}\label{eq5.7}
H_{x,t}(\hat{s})&=&\sup_{x(\cdot)\in B(x,t)}\, \sigma_{\varepsilon}^\dag
\left\lbrack\int_0^t\gamma(x(\tau),\tau)\, d\tau\right\rbrack
\sigma_{\varepsilon} \nonumber \\
&\ge& \sigma_{\varepsilon}^\dag
\left\lbrack\int_0^t\gamma(x_{\varepsilon}(\tau),\tau)\, d\tau\right\rbrack
\sigma_{\varepsilon}\ge \mu_{x,t} -\varepsilon .
\end{eqnarray}
Thus, along the direction of $\sigma_{\varepsilon}$, Lemma 2 and Equation\ (\ref{eq5.7}) yield
$$
\limsup_{\vert\vert s\vert\vert\rightarrow +\infty}\frac{\ln\left\vert {\cal E}(x,t;s)\right\vert^q}
{\vert\vert s\vert\vert^2}\ge q\lambda (\mu_{x,t} -\varepsilon) ,
$$
and for every $\lambda >(q\mu_{x,t} -q\varepsilon)^{-1}$,
$$
\limsup_{\vert\vert s\vert\vert\rightarrow +\infty}{\rm e}^{-\vert\vert s\vert\vert^2}
\vert {\cal E}(x,t;s)\vert^q =+\infty ,
$$
in contradiction with\ (\ref{eq5.6}). Therefore, $\lambda_{q}(x,t)\le (q\mu_{x,t} -q\varepsilon)^{-1}$. Taking $\varepsilon$ arbitrarily small one obtains $\lambda_{q}(x,t)\le (q\mu_{x,t} -0^+)^{-1}$, hence $\lambda_{q}(x,t)=(q\mu_{x,t})^{-1}$.
\paragraph*{}Finally, we always have $\overline{\lambda}_{q}(x,t)=1/q\mu_1\lbrack x(\cdot)=x\rbrack$ (see\ \cite{ADLM}) and $\mu_{x,t}\ge\mu_1\lbrack x(\cdot)=x\rbrack$ yields $\lambda_{q}(x,t)\le\overline{\lambda}_{q}(x,t)$, which completes the proof of the proposition.
%
%%%%%%%%%%%%%%%%%%%%
%
\section{Summary and perspectives}\label{sec6}
In this paper, we have studied the effects of propagation on the divergence of the solution to a linear amplifier driven by the square of a Gaussian field. We have considered a model in which the propagation is that of a free Schr\"odinger equation with a complex mass. For this model, we have explicitely determined the values of the coupling constant at which the moments of the solution diverge. We proved that the divergence yielded by a propagation-free calculation, i.e. in the limit of an infinite mass, cannot occur at a smaller coupling constant than the one obtained with a finite mass. This extends the results of Ref.\ \cite{ADLM} where such an inequality was proven in the diffusion case only, i.e. imaginary mass.
\paragraph*{}As explained in the conclusion of Ref.\ \cite{ADLM}, the stumbling block to going beyond the purely diffusive case was to control the growth of a complex Feynman path-integral. Our solution of this problem is based on the realization that, if $S$ is given by Eq.\ (\ref{eq2.1}), the Feynman path-integral can be rewritten as the Fourier integral of a distribution with compact support (Lemma 1). Control can then be obtained as a consequence of the Paley-Wiener theorem (Lemma 2).
\paragraph*{}In conclusion we outline some possible generalizations of this work. From a practical point of view, it would be interesting to find out whether there exists a class of $S$ of the form\ (\ref{eq2.1}) for which there is no propagation effects on the onset of the divergence, i.e. for which $\lambda_{q}(x,t)=\overline{\lambda}_{q}(x,t)$. In addition, since most Gaussian fields of physical interest admit Karhunen-Lo\`eve-type expansions, it would also be very interesting to find a way to generalize our solution to the case where the finite sum\ (\ref{eq2.1}) is replaced by an infinite sum. Other problems involve relaxing some of the assumptions in\ (\ref{eq1.1}). For instance, under what conditions on $S$ do the results carry over to the case where $\Lambda$ is replaced by ${\mathbb R}^d$ and ${\cal E}(x,0)\in L^2({\mathbb R}^d)$. It should also be checked whether our solution of the problem is robust with respect to the initial condition. If the answer is no, the size of the set of ${\cal E}(x,0)$ for which our results do not hold should be estimated according to physically relevant measures on the space of ${\cal E}(x,0)$.
\section*{Acknowledgments}
We warmly thank K. Yajima, C. Kopper, and G. Ben Arous for providing many valuable insights all along the completion of this work. The work of J. L. L. was supported by ASFOSR grant 49620-01-1-0154 and NSF grant DMR 01-279-26. J. L. L. also thanks the IHES at Bures-sur-Yvette, France, where part of this work was done.
%
%%%%%%%%%%%%%%%%%%%%
%
\appendix
\section{On the permutation of time integral and space supremum}\label{appA}
This appendix is devoted to the proof of the following lemma,
\paragraph*{}
\bigskip
\noindent {\bf Lemma A1.}
{\it Let $\Lambda$ be a compact pathwise connected metric space [with distance denoted by $d(\cdot ,\cdot)$], and $t>0$ a real number. Let $W$ be a real function on $\Lambda\times [0,t]$ such that $W(\cdot ,\tau)$ is continuous in $x$ uniformly in $\tau\in [0,t]$, and $\forall x\in\Lambda$, $W(x,\cdot)$ is piecewise continuous with a finite number of discontinuities. Then, for any $x\in\Lambda$
$$
\int_0^t\sup_{y\in\Lambda}W(y,\tau)\, d\tau =\sup_{x(\cdot)\in B(x,t)}\int_0^tW(x(\tau),\tau)\, d\tau ,
$$
where $B(x,t)$ is the set of continuous paths in $\Lambda$ satisfying $x(t)=x$.}
\paragraph*{}
\bigskip
\noindent {\it Proof.}
We obviously have
$$
\int_0^t\sup_{y\in\Lambda}W(y,\tau)\, d\tau\ge\sup_{x(\cdot)\in B(x,t)}\int_0^tW(x(\tau),\tau)\, d\tau ,
$$
and it remains to prove the inequality in the other direction. First we assume $W\in C^0(\Lambda\times [0,t])$. Since $\Lambda\times [0,t]$ is compact, $W$ is uniformly continuous, and for every $\epsilon >0$ we can find a number $\delta =\delta(\epsilon)>0$ such that if $\max\lbrace d(x,x'),\vert\tau -\tau'\vert\rbrace <\delta(\epsilon)$, then $\vert W(x,\tau)-W(x',\tau')\vert <\epsilon$. Moreover, $\delta(\epsilon)$ tends to zero with $\epsilon$. If $t/\delta(\epsilon)$ is not an integer, it is convenient to replace $\delta(\epsilon)$ by the smaller quantity $t/(1+[t/\delta(\epsilon)])$ where $[\cdot]$ denotes the integer part, and from now on we will assume that $t/\delta(\epsilon)$ is an integer. For a fixed $\epsilon >0$, let $N=N(\epsilon)=t/\delta(\epsilon)$, and let ${\cal R}_{\epsilon}$ be a finite partition of $\Lambda$ by sets of diameter at most $\delta(\epsilon)$. We observe that for any $0\le q\le N-1$ one has
\begin{equation}\label{eqa1}
\sup_{F\in {\cal R}_{\epsilon}}\sup_{0\le q\le N-1}
{\rm Osc}_{F\times [qt/N,\, (q+1)t/N]}W\le\epsilon ,
\end{equation}
where "${\rm Osc}$" denotes the oscillation of the function (namely, its sup minus its inf). We now choose once and for all a point $(x_{F,q},\tau_q)$ in each $F\times [qt/N,\, (q+1)t/N]$.
\paragraph*{}For each $0\le q\le N-1$ and any $\tau\in\lbrack q t/N,\, (q +1)t/N\rbrack$ one has
\begin{eqnarray*}
&&\sup_{y\in\Lambda}W(y,\tau)\le\sup_{F\in {\cal R}_\epsilon}
\sup_{F\times [qt/N,\, (q+1)t/N]}W \\
&&\le\sup_{F\in {\cal R}_\epsilon}\left\lbrack
\sup_{F\times [qt/N,\, (q+1)t/N]}W +W(x_{F,q},\tau_q)-\inf_{F\times [qt/N,\, (q+1)t/N]}W
\right\rbrack \\
&&\le\sup_{F\in {\cal R}_\epsilon}W(x_{F,q},\tau_q)
+\sup_{F\in {\cal R}_\epsilon}{\rm Osc}_{F\times [qt/N,\, (q+1)t/N]}W \\
&&\le\sup_{F\in {\cal R}_\epsilon}W(x_{F,q},\tau_q)
+\sup_{F\in {\cal R}_\epsilon}\sup_{0\le q\le N-1}{\rm Osc}_{F\times [qt/N,\, (q+1)t/N]}W \\
&&\le\sup_{F\in {\cal R}_\epsilon}W(x_{F,q},\tau_q)+\epsilon ,
\end{eqnarray*}
where we have used the inequality\ (\ref{eqa1}). Thus, choosing an atom $F_q\in {\cal R}_\epsilon$ such that $\sup_{F\in {\cal R}_\epsilon}W(x_{F,q},\tau_q)=W(x_{F_q,q},\tau_q)$, one has for any $\tau\in\lbrack qt/N,\, (q+1)t/N\rbrack$
\begin{equation}\label{eqa2}
\sup_{y\in\Lambda}W(y,\tau)\le W(x_{F_q,q},\tau_q)+\epsilon .
\end{equation}
Let $y_1$ and $y_2$ be two given points in $\Lambda$. We now define a continuous path $x_\epsilon$ in $\Lambda$ from $y_1$ to $y_2$. For any $1\le j\le N-1$ we choose a family of continuous paths $x_j$ from $\lbrack -t/2N^2,t/2N^2\rbrack$ to $\Lambda$ satisfying $x_j(-t/2N^2)=x_{F_{j-1},j-1}$ and $x_j(t/2N^2)=x_{F_j,j}$. We also choose a continuous path $x_0$ from $\lbrack 0,t/2N^2\rbrack$ to $\Lambda$ such that $x_0(0)=y_1$ and $x_0(t/2N^2)=x_{F_{0},0}$, and a continuous path $x_N$ from $\lbrack -t/2N^2,0\rbrack$ to $\Lambda$ such that $x_N(0)=y_2$ and $x_N(-t/2N^2)=x_{F_{N-1},N-1}$. The continuous path $x_\epsilon$ is defined by
\begin{equation}\label{eqa3}
x_\epsilon(\tau) =\left\lbrace
\begin{array}{ll}
x_0(\tau)&{\rm for}\ 0\le\tau\le t/2N^2, \\
x_{F_q,q}&{\rm for}\ qt/N+t/2N^2\le\tau\le (q+1)t/N-t/2N^2, \\
x_q(\tau -qt/N)&{\rm for}\ qt/N-t/2N^2\le\tau\le qt/N+t/2N^2\ {\rm for}\ q\ne 0, \\
x_N(\tau -t)&{\rm for}\ t-t/2N^2\le\tau\le t.
\end{array}
\right.
\end{equation}
One can observe that the Lebesgue measure of the time domain over which $x_\epsilon(\tau)\ne x_{F_q,q}$ is at most equal to $t/N$. Since $\Lambda$ is compact and $\forall\tau\in [0,t]$, $W(\cdot ,\tau)\in C^0(\Lambda)$, there is a finite number $M>0$ such that for any $(x,\tau)\in\Lambda\times [0,t]$ one has $\vert W(x,\tau)\vert\le M$. Thus, $\vert W(x_{F_q,q},\tau)-W(x_\epsilon(\tau),\tau)\vert\le 2M$ for any $\tau$ in $\lbrack qt/N,\, (q+1)t/N\rbrack$. Using the latter estimate, the remark below\ (\ref{eqa3}), and\ (\ref{eqa2}) one obtains
\begin{eqnarray}\label{eqa4}
\int_0^t\sup_{y\in\Lambda}W(y,\tau)\, d\tau &=&\sum_{q=0}^{N-1}\int_{qt/N}^{(q+1)t/N}
\sup_{y\in\Lambda}W(y,\tau)\, d\tau \nonumber \\
&\le&\epsilon t+\sum_{q=0}^{N-1}\int_{qt/N}^{(q+1)t/N}W(x_{F_q,q},\tau)\, d\tau \\
&\le& \epsilon t+\frac{2Mt}{N}+\int_0^tW(x_\epsilon(\tau),\tau)\, d\tau , \nonumber
\end{eqnarray}
\paragraph*{}Now, assume that there is a finite set of times independent of $x$, $\lbrace \tau_1,...,\tau_L\rbrace$, such that $\forall x\in\Lambda$ the set of times at which $W(x,\cdot)$ is discontinuous is a subset of $\lbrace\tau_1,...,\tau_L\rbrace$. Thus,\ (\ref{eqa4}) applies in each time interval $[\tau_i,\tau_{i+1}]$, $0\le i\le L$, with $\tau_0=0$ and $\tau_{L+1}=t$. Let $y_0,y_1,...,y_{L+1}$ be $L+2$ points in $\Lambda$ with $y_{L+1}=x$. Let $x_\epsilon$ be a continuous path in $\Lambda$ passing by $x_\epsilon(\tau_i)=y_i$ and defined by\ (\ref{eqa3}) in each time interval $[\tau_i,\tau_{i+1}]$. From Equation\ (\ref{eqa4}) in which one writes $C_i(\epsilon)=\epsilon +2M/N_i(\epsilon)$, it follows
\begin{eqnarray*}
\int_0^t\sup_{y\in\Lambda}W(y,\tau)\, d\tau &\le&
\sum_{i=0}^L\left\lbrack C_i(\epsilon)(\tau_{i+1}-\tau_i)+\int_{\tau_i}^{\tau_{i+1}}
W(x_\epsilon(\tau),\tau)\, d\tau\right\rbrack \\
&=&\sum_{i=0}^LC_i(\epsilon)(\tau_{i+1}-\tau_i) +\int_0^t W(x_\epsilon(\tau),\tau)\, d\tau \\
&\le&\sum_{i=0}^LC_i(\epsilon)(\tau_{i+1}-\tau_i) +\sup_{x(\cdot)\in B(x,t)}
\int_0^t W(x(\tau),\tau)\, d\tau ,
\end{eqnarray*}
where the last inequality results from the fact that $x_\epsilon\in B(x,t)$. The proof of Lemma A1 for the class of $W$ considered in this paragraph is completed by taking the limit $\epsilon\rightarrow 0$ and observing that the $C_i(\epsilon)$ tend to zero with $\epsilon$.
\paragraph*{}We are now ready to prove Lemma 1 in the general case. Since $W(\cdot ,\tau)$ is continuous in $x$ uniformly in $\tau\in [0,t]$, for any $\epsilon>0$, we can find $\delta=\delta(\epsilon)>0$ such that if $d(x,x')\le \delta$, then
$$
\sup_{\tau\in[0,t]}\vert W(x,\tau)-W(x',\tau)\vert\le\epsilon .
$$
Since $\Lambda$ is compact, we can find a finite covering by open balls of radius at most $\delta/2$, and therefore a finite partition of unity $\left(\chi_{k}\right)$ by continuous functions whose support has diameter at most $\delta$ (see \cite{Bou}, paragraph 4.3 in Chapter IX). For any $k$ we choose once for all a point $x_{k}\in{\rm supp}\chi_{k}$, and define the function
$$
W_{\epsilon}(x,\tau)=\sum_{k}W(x_{k},\tau)\chi_{k}(x) .
$$
For each fixed $\tau$, this function is obviously continuous in $x$, and for fixed $x$, it is piecewise continuous in $\tau$, with the possible discontinuity points belonging to a finite set which  can be chosen independent of $x$. From the previous result we have
$$
\int_{0}^{t}\sup_{y\in\Lambda}W_{\epsilon}(y,\tau)\, d\tau=
\sup_{x(\cdot)\in B(x,t)}\int_{0}^{t}W_{\epsilon}(x(\tau),\tau)\, d\tau .
$$
Since $\sup_{x\in\Lambda ,\tau\in[0,t]}\vert W(x,\tau)-W_{\epsilon}(x,\tau)\vert\le\epsilon$, we deduce that
\begin{eqnarray*}
\int_0^t\sup_{y\in\Lambda}W(y,\tau)\, d\tau&\le&\epsilon t+
\int_0^t\sup_{y\in\Lambda}W_{\epsilon}(y,\tau)\, d\tau \\
&=&\epsilon t+\sup_{x(\cdot)\in B(x,t)}\int_0^tW_{\epsilon}(x(\tau),\tau)\, d\tau \\
&\le& 2\epsilon t+\sup_{x(\cdot)\in B(x,t)}\int_0^tW(x(\tau),\tau)\, d\tau .
\end{eqnarray*}
Since the estimate holds for any $\epsilon>0$ the general result follows.
%
%%%%%%%%%%%%%%%%%%%
%
\section{Determination of the support of $\bm{g_{(x,T)}^{(m)}}$}\label{appB}
Let $g_{x,t}^{(m)}(u)$ be a distribution with compact support on ${\mathbb R}$ whose Fourier transform, $\Psi^{(m)}(x,t;\eta)\equiv({\cal F}g_{x,t}^{(m)})(\eta)$ with $\eta\in {\mathbb R}$, is the solution to\ (\ref{eq3.9}) with $V(x,t;\eta)=\eta U(x,t;\hat{s})$, where $U(x,t;\hat{s})=\sum_{i=1}^N\hat{k}(s)_i \varphi_i(x,t)$. This appendix is devoted to the determination of the support of $g_{(x,t)}^{(m)}$. We have modified the notation used in the text to make the dependence on $m$ explicit.
\paragraph*{}We begin with a technical lemma that will be useful in the sequel. Let $C_0^{\infty}({\mathbb R})$ denote the set of all smooth compactly supported functions in ${\mathbb R}$,
\paragraph*{}
\bigskip
\noindent {\bf Lemma B1.}
{\it For every $t>0$, $x\in\Lambda$, and $f\in C_0^{\infty}({\mathbb R})$, $\int_{{\mathbb R}} g_{x,t}^{(m)}(v)f(v)\, dv$ is an analytic function of $m$ on ${\mathbb C}^+ \equiv\lbrace m\in {\mathbb C}:\ {\rm Im}(m)>0\rbrace$, and $\int_{{\mathbb R}} g_{x,t}^{(m)}(v)f(v)\, dv=\lim_{\gamma\rightarrow 0^+} \int_{{\mathbb R}} g_{x,t}^{(m+i\gamma)}(v)f(v)\, dv$ for each real $m\ne 0$.}
\paragraph*{}
\bigskip
\noindent {\it Proof.}
As a Fourier transform of a function with compact supports on ${\mathbb R}$, $({\cal F}f)(\eta)$ is an analytic function of $\eta\in{\mathbb C}$. We have seen at the beginning of the proof of Lemma 1 that $\Psi^{(m)}(x,t;\eta)$, with $m\in {\mathbb C}^+$, is also analytic in $\eta$. Furthermore, if $\Lambda$ is a torus and $V$ is bounded on $\Lambda$ (which is the case), then (i) $\Psi^{(m)}(x,t;\eta)$ is analytic in $m\in {\mathbb C}^+$; and (ii) $\forall\eta\in {\mathbb C}$, $\Psi^{(m)}(x,t;\eta)=\lim_{\gamma\rightarrow 0^+}\Psi^{(m+i\gamma)}(x,t;\eta)$ for each real $m\ne 0$. We are indebted to K. Yajima for the proof of the latter result that we reproduce here for the sake of completeness\ \cite{Yajima}. Define
$$
{\cal U}_m(t)=\exp\left(\frac{it\Delta}{2m}\right),\ \ t\ge 0,\ \ m\in{\mathbb C}^+,
$$
and write the initial value problem\ (\ref{eq3.9}) in the the form of integral equation
$$
\Psi^{(m)}(t)=1-i\eta\int_0^t {\cal U}_m(t-\tau)V(\tau)\Psi^{(m)}(\tau)\, d\tau .
$$
Here $V(t)$ is the multiplication operator with $U(x,t;\hat{s})$. Let ${\cal B}$ denote the space of bounded operators in $L^2(\Lambda)$. By Fourier series expansion, it is evident that (a) $\vert\vert {\cal U}_m(t)\Psi^{(m)}(t)\vert\vert_2\le\vert\vert\Psi^{(m)}(t)\vert\vert_2$, viz. ${\cal U}_m(t)\in{\cal B}$ and $\vert\vert {\cal U}_m(t)\vert\vert\le 1$; (b) the function $\lbrack 0,\infty)\times\overline{{\mathbb C}^+}\ni (t,m)\rightarrow {\cal U}_m(t)\in{\cal B}$ is strongly continuous [viz. $(t,m)\rightarrow {\cal U}_m(t)f\in L^2(\Lambda)$ is continuous for every $f\in L^2(\Lambda)$]; and (c) for every $t\ge 0$, $m\rightarrow {\cal U}_m(t)\in{\cal B}$ is analytic for $m\in{\mathbb C}^+$ and $(d/dm){\cal U}_m(t)$ is norm continuous w.r.t. $(t,m)\in\lbrack 0,\infty)\times{\mathbb C}^+$. It follows from the boundedness of $V$ that the Dyson expansion\ \cite{RS}
\begin{eqnarray*}
&&D_m(t)={\cal U}_m(t)-i\eta\int_0^t{\cal U}_m(t-\tau)V(\tau){\cal U}_m(\tau)\, d\tau
+\cdots + \nonumber \\
&&(-i\eta)^n\int_{0<\tau_1<\cdots <\tau_n<t}{\cal U}_m(t-\tau_n)V(\tau_n)
\cdots V(\tau_1){\cal U}_m(\tau_1)\, d\tau_1\cdots d\tau_n +\cdots
\end{eqnarray*}
converges in the operator norm of ${\cal B}$ uniformly w.r.t. $(t,m)$ in every compact subset of $\lbrack 0,\infty)\times (\overline{{\mathbb C}^+}\backslash\lbrace 0\rbrace)$. Thus, the operator $D_m(t)$ enjoys the same properties (b) and (c) mentioned above as an operator valued function of $t$ and $m$. It is easy to check that $D_m(t)$ defines the propagator for\ (\ref{eq3.9}) and is unitary if $m$ is real. Hence the solution to\ (\ref{eq3.9}) satisfies the properties (i) and (ii).
\paragraph*{}Analyticity of $\Psi^{(m)}(x,t;\eta)$ and $({\cal F}f)(\eta)$ in $m\in {\mathbb C}^+$ and $\eta\in {\mathbb C}$ implies that
$$
\int_{{\mathbb R}} g_{x,t}^{(m)}(u)f(u)\, du\equiv
\int_{{\mathbb R}}\Psi^{(m)}(x,t;\eta)({\cal F}f)(-\eta)\frac{d\eta}{2\pi}
$$
is an analytic function of $m\in {\mathbb C}^+$. By Eq.\ (\ref{eq3.13}) and the fact that none of the constants $A$, $B$, and $C$ depend on $m$, $\vert\Psi^{(m)}(x,t;\eta)({\cal F}f)(-\eta)\vert$
is bounded by an integrable function of $\eta$ independent of $m$, from which it follows that the $m$-limit and the $\eta$-integral can be interchanged. Thus, according to (ii), one finds that for each real $m\ne 0$
\begin{eqnarray*}
\lim_{\gamma\rightarrow 0^+}\int_{{\mathbb R}} g_{x,t}^{(m+i\gamma)}(u)f(u)\, du
&=&\lim_{\gamma\rightarrow 0^+}\int_{{\mathbb R}}
\Psi^{(m+i\gamma)}(x,t;\eta)({\cal F}f)(-\eta)\frac{d\eta}{2\pi} \nonumber \\
&=&\int_{{\mathbb R}}\lim_{\gamma\rightarrow 0^+}\Psi^{(m+i\gamma)}(x,t;\eta)
({\cal F}f)(-\eta)\frac{d\eta}{2\pi} \\
&=&\int_{{\mathbb R}}\Psi^{(m)}(x,t;\eta)({\cal F}f)(-\eta)\frac{d\eta}{2\pi}
=\int_{{\mathbb R}} g_{x,T}^{(m)}(u)f(u)\, du , \nonumber
\end{eqnarray*}
which completes the proof of Lemma B1.
\paragraph*{}Let $a=\inf_{x(\cdot)\in B(x,t)}\int_0^tU(x(\tau),\tau;\hat{s})\, d\tau$ and $b=\sup_{x(\cdot)\in B(x,t)}\int_0^tU(x(\tau),\tau;\hat{s})\, d\tau =H_{x,t}(\hat{s})$. One has the following Lemma:
\paragraph*{}
\bigskip
\noindent {\bf Lemma B2.} {\it For every $t>0$, $x\in\Lambda$, and $m\in\overline{{\mathbb C}^+}\backslash\lbrace 0\rbrace$, the support of $g_{x,t}^{(m)}$ is {\it equal} to $\lbrack a,b\rbrack$.}
\paragraph*{}
\bigskip
\noindent {\it Proof.}
First, consider the case $m=i\gamma$, $\gamma >0$. Denote by $\alpha\lbrack x(\cdot)\rbrack$ the functional $\alpha\lbrack x(\cdot)\rbrack\equiv\int_0^tU(x(\tau),\tau;\hat{s})\, d\tau$. By boundedness of $\nabla V$ over $\Lambda\times\lbrack 0,t\rbrack$, $\exists A>0$ such that $\forall x,y\in\Lambda$ and $0\le\tau\le t$,
$$
\vert U(x,\tau;\hat{s})-U(y,\tau;\hat{s})\vert =\left\vert\int_{\overline{x,y}}\nabla U(x',\tau;\hat{s})\cdot dx'\right\vert
\le A\| x-y\| ,
$$
where $\overline{x,y}$ denotes the segment of geodesic from $x$ to $y$ and $\|\cdot\|$ is the usual Euclidean distance. This implies that $\forall x(\cdot),y(\cdot)\in B(x,t)$,
\begin{eqnarray*}
\vert\alpha\lbrack x(\cdot)\rbrack -\alpha\lbrack y(\cdot)\rbrack\vert &\le&
\int_0^t \vert U(x(\tau),\tau;\hat{s})-U(y(\tau),\tau;\hat{s})\vert\, d\tau \\
&\le&A\int_0^t \| x(\tau)-y(\tau)\|\, d\tau \\
&\le& At\sup_{0\le\tau\le t}\| x(\tau)-y(\tau)\| ,
\end{eqnarray*}
which shows that $\alpha\lbrack x(\cdot)\rbrack$ is a continuous functional of $x(\cdot)\in B(x,t)$ with the uniform norm on $\lbrack 0,t\rbrack$. Let $h\in C_0^{\infty}({\mathbb R})$ a real positive test function with support in $\lbrack a,b\rbrack$ and $\sup_{u\in {\mathbb R}}h(u)=1$. From the continuity of $\alpha\lbrack x(\cdot)\rbrack$ it follows that $\exists x_0(\cdot)\in B(x,t)$ such that $h(\alpha\lbrack x_0(\cdot)\rbrack)=1$. By continuity of $h$ and $\alpha\lbrack x(\cdot)\rbrack$ it follows that $\forall\varepsilon >0$, $\exists\delta >0$ such that $\vert h(\alpha\lbrack x(\cdot)\rbrack)-1\vert <\varepsilon$ for every $x(\cdot)\in B_0(\delta)\equiv\lbrace x(\cdot)\in B(x,t):\ \sup_{0\le\tau\le t}\| x(\tau)-x_0(\tau)\| <\delta\rbrace$. Take $\varepsilon =1/2$, in this case $h(\alpha\lbrack x(\cdot)\rbrack)>1/2$ for every $x(\cdot)\in B_0(\delta)$ and one has
\begin{eqnarray*}
\int_{{\mathbb R}} g_{x,t}^{(i\gamma)}(u)h(u)\, du &=&
\int_{x(\cdot)\in B(x,t)}{\rm e}^{-\frac{\gamma}{2}
\int_0^t\dot{x}(\tau)^2d\tau} h(\alpha\lbrack x(\cdot)\rbrack)\,
d\lbrack x(\cdot)\rbrack \nonumber \\
&\ge&\int_{x(\cdot)\in B_0(\delta)}
{\rm e}^{-\frac{\gamma}{2}
\int_0^t\dot{x}(\tau)^2d\tau} h(\alpha\lbrack x(\cdot)\rbrack)\,
d\lbrack x(\cdot)\rbrack \\
&>&\frac{1}{2}\int_{x(\cdot)\in B_0(\delta)}
{\rm e}^{-\frac{\gamma}{2}\int_0^t\dot{x}(\tau)^2d\tau}
d\lbrack x(\cdot)\rbrack . \nonumber
\end{eqnarray*}
Since the set of the Brownian paths $x(\cdot)$ that are in $B_0(\delta)$ has a strictly positive Wiener measure, the last term is strictly positive and one finds
\begin{equation}\label{eqb1}
\int_{{\mathbb R}} g_{x,t}^{(i\gamma)}(u)h(u)\, du >0.
\end{equation}
If there was an open subset of $\lbrack a,b\rbrack$ not intersecting the support of $g_{x,t}^{(i\gamma)}$, it would be possible to choose the support of $h$ outside the one of $g_{x,t}^{(i\gamma)}$, yielding $\int g_{x,t}^{(i\gamma)}(u)h(u)\, du=0$ in contradiction with Eq.\ (\ref{eqb1}). Thus, for every $x\in\Lambda$ and $\gamma >0$, the support of $g_{x,t}^{(i\gamma)}$ is equal to $\lbrack a,b\rbrack$.
\paragraph*{}Consider now the general case $m\in\overline{{\mathbb C}^+}\backslash\lbrace 0\rbrace$. Assume that there is an open subset of $\lbrack a,b\rbrack$ not intersecting the support of $g_{x,t}^{(m)}$. In this case it is possible to choose the support of $h$ outside the one of $g_{x,t}^{(m)}$, yielding $\int g_{x,t}^{(m)}(u)h(u)\, du =0$. By Lemma B1 the support of $g_{x,t}^{(z)}$ must vary continuously with $z\in\overline{{\mathbb C}^+}$, whence the support of $h$ can be taken small enough such that there is a open subset ${\cal V}(m)\subset {\mathbb C}^+$ with $m\in\overline{{\cal V}(m)}$ and $\int g_{x,t}^{(z)}(u)h(u)\, du=0$ identically in ${\cal V}(m)$. From the analyticity of $\int g_{x,t}^{(z)}(u)h(u)\, du$ in $z$ on ${\mathbb C}^+$ (Lemma B1), it follows immediately that $\int g_{x,t}^{(z)}(u)h(u)\, du=0$ identically in all ${\mathbb C}^+$, in contradiction with Eq.\ (\ref{eqb1}), which completes the proof of Lemma B2. In particular, one has $\sup\lbrace v:v\in {\rm supp}g_{x,t}^{(m)}\rbrace =b=H_{x,t}(\hat{s})$, which is the result used in the proof of Lemma 2.
%
%%%%%%%%%%%%%%%%%%%%
%

%
%
\end{document}